\documentstyle[11pt,epsfig,float]{article}
\begin{document}
\begin{center}
{\bf{Dynamical Matrices and Interatomic-Force Constants from Wave-Commensurate Supercells}}

\vspace*{0.50in}

Hadley M. Lawler\\

{\em{Department of Physics, University of Maryland}}\\
{\em{College Park, Maryland 20742-4111}}

\vspace*{0.25in}

Eric K. Chang\\

{\em{University of Modena, Via Universita 4, 41100 Modena, Italy}}

\vspace*{0.25in}

Eric L. Shirley\\
{\em{Optical Technology Division}} 
{\em{National Institute of Standards and Technology}}\\
{\em{100 Bureau Dr., Stop 8441, Gaithersburg, MD 20899-8441}}
\end{center}

\vspace*{0.25in}
\begin{center}
\Large{Abstract}
\end{center}
\indent We apply standard, first-principles calculations to a complete treatment of lattice dynamics in the harmonic approximation.  The algorithm makes use of the straightforward ``frozen-phonon'' approach to the calculation of vibrational spectra and addresses some limitations of the method.  Our prescription's validity is independent of crystal structure.  It treats polar crystals in a general way, and it incorporates interatomic-force constants reaching beyond the supercell that is considered.  For a range of materials, phonon dispersion exhibits the close agreement with experiment that is now characteristic of first-principles schemes.  Results for graphite, Si and GaAs are presented.  
\newpage

\begin{center}
\Large{I. Introduction}
\end{center}

\indent  Several years ago, density-functional theory (DFT)~\cite{DFT} opened the way to detailed knowledge of phonon spectra from first principles.  The two most common approaches today are density-functional perturbation theory (DFPT)~\cite{DFPT}, where linear corrections to the equilibrium electronic wave functions are calculated with respect to some perturbation, in this case ionic displacements, and the ``frozen-phonon'' method, which solves for the ground state with the displacements ``frozen'' in~\cite{KuncinFrank}.  The latter is taken up here.

\indent  Direct-method calculations, in conjunction with the Hellman-Feynman force theorem~\cite{Hf}, have evaluated restoring forces and phonon frequencies by employing two different types of supercells.  Initial work utilized wave-commensurate supercells, which map the phonon's periodicity to the supercell.  This allowed for the evaluation of dynamical matrices for a very small set of wave vectors where the phonon polarizations were known \textit{a priori}~\cite{apriori}.  Later, supercells were designed to compute interplanar forces.  Transforming the planar forces into the real-space interatomic-force constants (IFCs) made possible the expansion of the dynamical matrix throughout the Brillouin zone (BZ)~\cite{WC92}.  Our strategy returns to wave-commensurate supercells, but displaces ions along Cartesian rather than along normal coordinates.

\indent  In this respect, our method resembles that of DFPT, where the dynamical matrices are derived directly in momentum space, and then interpolated~\cite{bologna}.  Accordingly, similar techniques can be exploited to account for the longitudinal-transverse (LO-TO) optical phonon splitting, which are somewhat different than those often combined with frozen-phonon calculations~\cite{WC92,Det98,Par98,Par00}.

\indent  The present scheme, at the level of DFT, exploits the periodicity of the supercells and includes all the interactions, while requiring no prior knowledge of phonon modes.  The computation of planar-force constants discounts interactions extending beyond the supercell, even if they are numerically represented in a calculation~\cite{Ack97,Frank95}.    This may suggest the possibility of a full dispersion calculation with smaller supercells than have been necessary until now.

\indent The method described here proceeds in two steps:  first, a set of high-symmetry wave vectors, $\lbrace{\bf k}\rbrace$, is chosen, and the corresponding reciprocal-space, interatomic-force constant matrices $\lbrace\tilde{\bf{C}}(\bf {k})\rbrace$ are computed; and second, the counterpart real-space matrices are extracted through a fit.  A fit is performed, rather than the familiar Fourier transform, to conveniently accomodate an irregular wave-vector grid while abiding by crystal symmetry.  This distinction from the interpolation in Ref.~\cite{bologna} reflects the differing requirements of each treatment.  Within DFPT, the phonon and electron grids must be equivalent and are typically regular, whereas in the present treatment there is freedom to select the wave-vector grid, provided that the corresponding periodicities are commensurate with carefully chosen supercells.  

\indent  If, at the Bravais lattice site indicated by the vector ${\bf R}$, 
the $a^{th}$ coordinate of the ${\tau}^{th}$ ion's position is given by:
\begin{equation}
r_{{\tau}{a}}({\bf{R}})=R_{a}+s_{{\tau}{a}}+{x}_{{\tau}{a}}({\bf{R}}),
\label{position}
\end{equation}
where the first two terms on the right sum to the equilibrium position, then the real-space IFCs are defined by:
\begin{equation}
C_{ {\tau}{\tau}^{'}{a}{b}}({\bf{R}})=
\frac{{\partial}^{2}V}{ {\partial}{x}_{{\tau}{a}}({\bf{R}}^{'})
{\partial}{x}_{{\tau}^{'}b}(\bf{R}+{\bf{R}}^{'})},
\label{realspace}
\end{equation}
where $V$ is the Born-Oppenheimer energy surface.
The translational symmetry of the crystal requires that the IFCs are independent of ${\bf R}'$.

\indent The real-space and reciprocal-space matrices are related by:
\begin{equation}
{\tilde{\bf{C}}({\bf{k}}})={\sum}_{{\bf{R}}}{\bf{C}}({\bf{R}})e^{i{\bf{k}}{\cdot}{\bf{R}}}.
\label{identity}
\end{equation}
The above expression establishes the general fit condition on the real-space matrices, and gives an expression for the reciprocal-space matrices at arbitrary $\bf{k}$ once the real-space matrices are known.  

\indent The dynamical matrix is obtained by dividing the reciprocal-space IFCs by the ionic masses:
\begin{equation}
{D}_{ {\tau}{\tau}^{'}{a}{b}}({\bf{k}})=
\frac{1}{\sqrt{m_{\tau}m_{\tau'}}}\tilde{C}_{ {\tau}{\tau}^{'}{a}{b} }({\bf{k}}).
\label{DM}
\end{equation}
\indent  Polar crystals, or those in which sub-lattice shifts are accompanied by macroscopic electric dipoles, exhibit splitting of the small-momentum, LO-TO phonon degeneracy.  This effect requires additional consideration to be incorporated into the above procedure.  In this work, such effects are incorporated with a method analogous to that used in earlier work, where a distinction is made between ``covalent'' and dipole-dipole contributions to the real-space IFC matrices~\cite{bologna,LandG}.

\begin{center}\Large{II.  Sampling of the Brillouin Zone}
\end{center}

\indent  The periodic ionic displacements entering a frozen-phonon calculation are constructed with cells larger than the primitive cell, or supercells.  In particular, cells are constructed such that ionic positions are given by
\begin{equation}r_{{\tau}{a}}({\bf{R}})=
R_{a}+s_{{\tau}a}+{d}_{{\tau}{a}}\cos(\bf{k{\cdot}R}).
\label{per}
\end{equation} 
Above, the displacement from equilibrium in Eq.~\ref{position}, 
$x_{{\tau}{a}}(\bf{R})$,
is written to represent a standing wave of wave vector ${\bf k}$.  As an illustration, considering constructions four times the size of the primitive cell or smaller, the face-centered cubic lattice can be generated by supercells corresponding to inequivalent wave vectors along the following Cartesian directions:  
$(1,0,0)$, $(1,1,1)$, $(1,1,0)$, $(2,1,0)$ and $(3,3,1).$  Lower-symmetry wave-vectors can be represented with larger supercells.
The Hellman-Feynman force is computed as
\begin{displaymath}
F_{{\tau}{a}}({\bf{R}})=
-\frac{{\partial}{V}}{\partial{r}_{{\tau}{a}}({\bf{R}})}
\end{displaymath} 
within a DFT calculation~\cite{YandC}.  Electronic states are solved  using pseudopotential techniques~\cite{Hamman,Vanderbilt,KandB}, specially-chosen points to sample the BZ~\cite{Chadi}, and well-known methods for handling large matrices~\cite{Teter}. 

\indent The harmonic approximation implies the forces and IFCs are related by:
\begin{eqnarray}
F_{{\tau}{a}}({\bf{R}}^{''})&=&-
{\sum}_{{\bf{R}}^{'}{\tau}^{'}{b}}
{C}_{{\tau}{\tau}^{'}{a}{b}}({\bf{R}}^{'}-{\bf{R}}^{''}){x}_{{\tau}^{'}{b}}({\bf{R}}^{'})\\
\nonumber
&=&-
{\sum}_{{\bf{R}}^{'}{\tau}^{'}b}{C}_{{\tau}{\tau}^{'}ab}({\bf{R}}^{'})d_{{\tau}^{'}b}\cos({\bf{k{\cdot}(R^{''}+{R}^{'})}}),
\end{eqnarray}
where ${\bf R}^{'}$ is shifted, and the ionic displacements from equilibrium are substituted from Eq.~\ref{per}.
If $d_{{\tau}^{'}{b}}$ is nonzero for only a single ion $\tau^{'}$ and direction $b$, and is varied over two successive DFT calculations, the following demonstrates that the finite difference in Hellman-Feynman forces can be related directly to the dynamical matrix.  Starting with
\begin{equation}
\frac{{\partial}F_{{\tau}a}({\bf{R}}^{''})}{{\partial}d_{{\tau}^{'}b}}=
-
\sum_{{\bf{R}}^{'}}{C}_{{\tau}{\tau}^{'}{a}{b}}({\bf{R}}^{'})\cos({\bf{k{\cdot}(R^{''}+{R}^{'})}}),
\end{equation} 
and using the basic identity
\begin{displaymath}
\cos(a+b)=\cos(a)\cos(b)-\sin(a)\sin(b)
\end{displaymath}
 and Eq.~\ref{identity}, we obtain
\begin{eqnarray}
{\mathrm{Re}}~(\tilde{{C}}_{{\tau}{\tau}^{'}{a}{b}}({\bf{k}}))=
-\frac{1}{2\cos({\bf{k}\cdot{\bf{R}}^{''}})}
\left(
\frac{{\partial}F_{{\tau}a}({\bf{R}}^{''})}{{\partial}d_{{\tau}^{'}b}}
+\frac{{\partial}F_{{\tau}a}({-\bf{R}}^{''})}{{\partial}d_{{\tau}^{'}b}}
\right)\\\nonumber
{\mathrm{Im}}~(\tilde{{C}}_{{\tau}{\tau}^{'}{a}{b}}({\bf{k}}))=
\frac{1}{2\sin({\bf{k}\cdot{\bf{R}}^{''}})}
\left(
\frac{{\partial}F_{{\tau}a}({\bf{R}}^{''})}{{\partial}d_{{\tau}^{'}b}}
-\frac{{\partial}F_{{\tau}a}({-\bf{R}}^{''})}{{\partial}d_{{\tau}^{'}b}}
\right).
\label{partials}
\end{eqnarray}
\noindent The lattice vector ${\bf{R}}^{''}$ is arbitrary, but chosen within the supercell such that the denominator is nonzero.  In the case that ${\bf k}$ is along certain high-symmetry directions and is half of a reciprocal-lattice vector, ${\bf{k}\cdot{\bf{R}}^{''}}$ is a multiple of $\pi$ for all ${\bf R}^{''}$.  At these points the condition $\tilde{\bf{C}}({\bf{k}})=\tilde{\bf{C}}(\bf{-k})$ is valid, and by Eq.~\ref{identity}, the equality $\tilde{\bf{C}}({\bf{k}})=\left(\tilde{\bf{C}}(\bf{-k})\right)^{*}$ follows, demonstrating that the reciprocal-space tensor must be purely real at any such point.

\indent The practical result of Eq.~8 is that an entire row of the reciprocal-space matrix is known from two calculations, each equivalent to a supercell calculation of total energy.  Typically, knowledge of the reciprocal-space IFCs at two or three wave vectors along each high-symmetry direction is sufficient for a good phonon dispersion throughout the BZ.  
Although complications arise from the fitting to a real-space representation, and the interpolated dynamical matrices do depend on real-space cutoffs, at least at the level of DFT, the formalism here incoporates interactions between ions that are not local to a supercell.  While the standard approximations to the electronic ground state are present in the DFT-calculated dynamical matrices, through construction of supercells representing standing waves mapped to the BZ point considered, Eq.~\ref{partials} formally accounts for all the IFCs.

\indent  In order to carry out the interpolation of the dynamical matrix, a discrete Fourier transform on the complete set of vectors within a parallelopiped box may appear tempting.  If one wishes to respect the symmetry of the crystal, however, this tactic will not work.  Generally, the set of vectors spanned by such a box will include some vectors within a symmetry-related subset, and not others.  If the set is inclusive enough, the artificial asymmetry introduced may be negligible, because ${\bf{C}}({\bf{R}})$ is expected to diminish reasonably quickly with ${\bf{R}}$, and the set can include all members of any subset whose components have significant magnitude.

\indent  Rather than performing a Fourier transform, by overdetermining the real-space IFC matrices used in Eq.~\ref{identity} and then fitting them, the crystal symmetry is preserved here, and no nuisance is posed by the choice of wave-vector sampling.  The following section details how this procedure is implemented.  Key features of the formulation are that all symmetries of the crystal are respected, the symmetries are exploited to minimize the number of parameters to be fit, and frequencies of the acoustic phonons are guaranteed to go to zero in the limit of small wave vector.

\indent  The next section assumes that the real-space matrices are short-ranged, and do not include the long-range Coulombic, ion-ion interactions.  As a result, the treatment, without being supplemented, can be applied only to nonpolar crystals.  A separate section subsequently extracts the contributions from the Coulombic interactions and extends the formalism to polar crystals. 

\begin{center}
\Large{III. Fitting Algorithm}
\end{center}  
\indent  The Born-Oppenheimer energy surface of a lattice is given by 
$V(\lbrace {\bf x}_{\tau}({\bf R}) \rbrace)$,
where ${\bf x}_{\tau}({\bf R})$ denotes the 
displacement from equilibrium of the atom located
at ${\bf R} + {\bf s}_{\tau}$, or at basis vector 
${\bf s}_{\tau}$ indexed by $\tau$ in the cell specified by
the lattice vector ${\bf R}$. The matrices,
$\lbrace {\bf C}_{\tau\tau'}({\bf R}) \rbrace$, are the $3 \times 3$ 
matrices denoting the IFCs. They satisfy
\[
V(\lbrace {\bf x}_{\tau}({\bf R}) \rbrace) 
= V_0 + {1 \over 2} \sum_{{\bf S,R},\tau \tau'}
{{\bf x}_{\tau}}^T({\bf S}) \cdot 
{\bf C}_{\tau \tau'}({\bf R}) \cdot
{\bf x}_{\tau'}({\bf S+R}) + \cdots.
\]
Because of crystal symmetry,
there are simple symmetry relationships between interatomic-force constants. Let
${\bf U}_n$ be a symmetry operator of the crystal group (i.e., a
$3 \times 3$ proper or improper rotation matrix) indexed by $n$,
which satisfies,

\begin{equation}
{\bf U}_n {\bf s}_{\tau} = 
{\bf s}_{\sigma} + {\bf g}_{n} + {\bf R}({\bf U}_n,{\bf s}_{\tau})
\label{b1}
\end{equation}

\begin{equation}
{\bf U}_n {\bf s}_{\tau'} = 
{\bf s}_{\sigma'} + {\bf g}_{n} + {\bf R}({\bf U}_n,{\bf s}_{\tau'}).
\label{b2}
\end{equation}

\begin{equation}
{\bf R'}={\bf U}_n{\bf R} +
{\bf R}({\bf U}_n,{\bf s}_{\tau'})-{\bf R}({\bf U}_n,{\bf s}_{\tau}),
\label{b3}
\end{equation}
where ${\bf g}_{n}$ is the glide associated
with ${\bf U}_n$, ${\bf s}_{\tau}$ is the atomic position
of the $\tau^{th}$ atom relative to a Bravais lattice site and,
${\bf R}({\bf U}_n,{\bf s}_{\tau})$, ${\bf R'}$, and ${\bf R}$
are lattice vectors. The basic symmetry transformation
relating interatomic force constants  is
\begin{equation}
({\bf U}_n {\bf e}_a)^T \cdot {\bf C}_{\sigma \sigma'}({\bf R'})
\cdot ({\bf U}_n {\bf e}_b) =
{\bf e}_a^T \cdot {\bf C}_{\tau \tau'}({\bf R}) \cdot {\bf e}_b,
\label{basic}
\end{equation}
where ${\bf e}_a$ and ${\bf e}_b$ are unit-vectors in Cartesian
directions labeled by the indices $a$ and $b$.

We may call the triplets $({\bf R},\tau,\tau')$ and
$({\bf R'},\sigma,\sigma')$ symmetry equivalent if they satisfy 
Eqs.~\ref{b1}, \ref{b2}, and \ref{b3}, for some
symmetry operation of the crystal group, ${\bf U}_n$.
We may divide these triplets into triplet sets in which all
of the members
within the same set are symmetry equivalent. 
After we divide all the triplets into sets,
we assign
to each triplet an ordered pair, $(j,i)$, 
where $j$ labels the set to which
the triplet belongs, and $i$ labels 
which member in the set the triplet corresponds to. There is then a 
one-to-one correspondence between the triplet $({\bf R},\tau,\tau')$ and the
ordered pair $(j,i)$. 


\indent  We are interested in computing the general
form of ${\bf C}_{\tau\tau'}({\bf R})$, that is, 
finding a minimal set of $3 \times 3$ basis matrices, 
$\lbrace {\bf F}_k,k=1,\ldots,M \rbrace$, in terms of which  
${\bf C}_{\tau\tau'}({\bf R})$ can be generally expressed
as a linear combination. 
That is, we have
\begin{equation}
{\bf C}_{\tau\tau'}({\bf R}) = \sum_{k=1}^M 
 a_k {\bf F}_k.
\end{equation}
To find  $\lbrace {\bf F}_k,k=1,\ldots,M \rbrace$,
we use the basic symmetry relation Eq.~\ref{basic}, and other
elementary properties of the  matrix ${\bf C}_{\tau\tau'}({\bf R})$,
to form general constraints on ${\bf C}_{\tau\tau'}({\bf R})$. We
express these constraints as homogeneous 
linear equations that give us $\lbrace {\bf F}_k \rbrace$.
Plainly
we have
\begin{equation}
{{\partial^2 V} \over 
{\partial x_{{\tau'}b}({\bf R})  \partial x_{{\tau}a}({\bf 0}) }} 
 = 
{{\partial^2 V} \over 
{\partial x_{{\tau}a}({\bf 0})  \partial x_{{\tau'}b}({\bf R}) }},
\label{mixed}
\end{equation}
and
\begin{equation}
{{\partial^2 V} \over 
{\partial x_{{\tau}a}({\bf S})  \partial x_{{\tau'}b}({\bf R +S}) }} 
 = {{\partial^2 V} \over 
{\partial x_{{\tau}a}({\bf 0})  \partial x_{{\tau'}b}({\bf R}) }}
\label{trans}. 
\end{equation}
Eq.~\ref{mixed} is the equality of mixed partials, and 
Eq.~\ref{trans} results from the discrete translational symmetry
of the crystal. Combining Eq.~\ref{mixed} and Eq.~\ref{trans} and
setting ${\bf S} = - {\bf R}$, we obtain
\begin{equation}
{{\partial^2 V} \over 
{\partial x_{{\tau'}b}({\bf 0})  \partial x_{{\tau}a}({\bf -R}) }} 
 = 
{{\partial^2 V} \over 
{\partial x_{{\tau}a}({\bf 0})  \partial x_{{\tau'}b}({\bf R}) }} .
\label{partial}
\end{equation}
Combining the definition in Eq.~\ref{realspace} and Eq.~\ref{partial}, we 
have
\begin{equation}
{\bf C}^T_{\tau'\tau}({\bf -R}) =
{\bf C}_{\tau\tau'}({\bf R}).
\label{partial1}
\end{equation}
From this it follows that $\tilde{{\bf C}}_{\tau\tau'}({\bf k})$
is a Hermitian matrix, if taken as a 
$3 N \times3 N $ matrix, where $N$ is the number of atoms
in the unit cell.
Because Eq.~\ref{partial1} is equivalent to the Hermiticity of the
reciprocal-space IFC $\tilde{{\bf C}}_{\tau\tau'}({\bf k})$, we will refer to
Eq.~\ref{partial1} as the ``Hermiticity condition.''

For a given triplet 
$({\bf R},\tau,\tau')$, 
we define the {\it little group} and {\it transpose set}
($L_{{\bf R},\tau,\tau'}$ and $T_{{\bf R},\tau,\tau'}$) as follows.
$L_{{\bf R},\tau,\tau'}$ is the subgroup of
the crystal group that transforms the triplet $({\bf R},\tau,\tau')$ into
itself. $T_{{\bf R},\tau,\tau'}$ is the subset of
 the crystal group that transforms the triplet
$({\bf R},\tau,\tau')$ to $(- {\bf R},\tau',\tau)$.
$T_{{\bf R},\tau,\tau'}$ may or may not be a group.
From Eq.~\ref{basic}, we have
\begin{equation}
{\bf C}_{\tau' \tau}(-{\bf R}) = {\bf U}_n^T {\bf C}_{\tau \tau'}({\bf R}) {\bf U}_n,
 {\bf U}_n \in T_{{\bf R},\tau,\tau'}.
\label{intermediate}
\end{equation}
Combining Eq.~\ref{intermediate} with 
the Hermiticity condition (Eq.~\ref{partial1}), we have
\begin{equation}
{\bf C}^T_{\tau \tau'}({\bf R}) = {\bf U}_n^T {\bf C}_{\tau \tau'}({\bf R}) {\bf U}_n,
 {\bf U}_n \in T_{{\bf R},\tau,\tau'}.
\label{constraint-transpose}
\end{equation}
This constrains the transpose of the real-space IFCs. 
In particular, if one has cubic
symmetry, it implies that the diagonal blocks
of the dynamical matrix must be  symmetric. Further constraints
can be imposed 
by the little group:
\begin{equation}
{\bf C}_{\tau \tau'}({\bf R}) = {\bf U}_n^T {\bf C}_{\tau \tau'}({\bf R}) {\bf U},
 {\bf U}_n \in L_{{\bf R},\tau,\tau'}.
\label{constraint-little}
\end{equation}
Eq.~\ref{constraint-transpose} defines
the {\it transpose-set constraints}, and
Eq.~\ref{constraint-little} defines the
{\it little-group constraints}. The next step is to exploit the above constraints toward finding the basis matrices, $\lbrace {\bf F}_k\rbrace$.  Later, further constraints are introduced to enforce the acoustic sum rule.

We may define another set of 
$3 \times 3$ basic matrices, 
$\lbrace {\bf B}_i, i=1,2,\ldots,9 \rbrace$, as,
\[
{\bf B}_1 = {\bf \hat{x} \hat{x}},
{\bf B}_2 = {\bf \hat{y} \hat{y}},
{\bf B}_3 = {\bf \hat{z} \hat{z}},
\]

\[
{\bf B}_4 = {\bf \hat{y} \hat{z}},
{\bf B}_5 = {\bf \hat{x} \hat{z}},
{\bf B}_6 = {\bf \hat{x} \hat{y}},
\]

\[
{\bf B}_7 = {\bf \hat{z} \hat{y}},
{\bf B}_8 = {\bf \hat{z} \hat{x}},
{\bf B}_9 = {\bf \hat{y} \hat{x}}.
\]

Here we employ the dyadic notation whereby ${\bf \hat{z} \hat{y}}$
denotes a $3 \times 3$ matrix with all zero entries except for
a 1 in the 3rd row and 2nd column.
We may define the scalar product of two 
$3 \times 3$ matrices in terms of the basic matrices,
\begin{equation}
\langle {\bf B}_i , {\bf B}_j \rangle = \delta_{ij}.
\end{equation}
We define the entries of the $9 \times 9$ matrices, ${\bf S}$, and
${\bf M}(n)$, by,
\begin{equation}
S_{ij} = \langle {\bf B}_i , {\bf B}_j^T \rangle 
\end{equation}
and
\begin{equation}
M_{ij}(n) = \langle {\bf B}_i , {\bf U}_n^T {\bf B}_j {\bf U}_n \rangle. 
\end{equation}
 If we express
 ${\bf C}_{\tau \tau'}({\bf R})$ as a linear combination of the nine basic
matrices, i.e., as
\begin{equation}
{\bf C}_{\tau \tau'}({\bf R}) = \sum_{i=1}^{9} \alpha_i {\bf B}_i,
\end{equation}
the little-group constraints (Eq.~\ref{constraint-little})
become
\begin{equation}
\sum_{j=1}^{9} (M_{ij}(n) - \delta_{ij}) \alpha_j = 0,
{\bf U}_n \in L_{{\bf R},\tau,\tau'},
\label{one}
\end{equation}
and the transpose-set constraints (Eq.~\ref{constraint-transpose}),
\begin{equation}
\sum_{j=1}^{9} (M_{ij}(n) - S_{ij}) \alpha_j = 0,
{\bf U}_n \in T_{{\bf R},\tau,\tau'}.
\label{two}
\end{equation}
If there are $N_L$  operations in the little group and
$N_T$ operations in the transpose set, there
are a total of $N' = 9(N_L+N_T)$ 
constraint equations, not all of
which are independent. Let us define a set of $N'$ 9-component vectors,
$\lbrace {\bf \beta}^{(n)},n=1,\ldots,N' \rbrace$, 
such that the system of
equations,
\begin{equation}
\sum_{i=1}^{9} \beta^{(n)}_i \alpha_i = 0, n=1,\ldots,N',
\end{equation}
is equivalent to the combined system of Eq.~\ref{one} and Eq.~\ref{two}.
We then define 
$\lbrace {\bf v}^{(n)}, n= 1,\ldots,M' \rbrace$, with 
$M' \leq N'$, as a Gramm-Schmidt,
orthonormal basis that spans the same space as  
$\lbrace \beta^{(n)} \rbrace$.
We define a $9 \times 9$ dimensional projector matrix,
${\bf P}$,
using
\begin{equation}
P_{ij} = \sum_{n=1}^{M'} {v}^{(n)}_i {v}^{(n)}_j.
\end{equation}
Let $\lbrace {\bf w}^{(k)},k=1,\ldots,M \rbrace$ be
the complete set of $M$ orthonormal eigenvectors
of ${\bf P}$ with zero eigenvalue. 
We then have
\begin{equation}
{\bf F}_k = \sum_{i=1}^9 w^{(k)}_i {\bf B}_i.
\end{equation}

\indent  In what follows, we consistently use the following symbols to denote
the various different kinds of indices involved. All of the below indices
are integers.
\begin{itemize}
\item $\mu$ : indexes the ${\bf k}_\mu$-point in the 
            BZ.
\item $a,b$ : are Cartesian vector indices.
\item $j$ : refers to a set of symmetry-equivalent dynamical matrices in real space.
\item $i$ : refers to a member in a set.
\item $k$ : refers to the basis matrices of which the IFCs
            in real-space are a linear combination.
\item $\sigma,\sigma'$ : refer to the atomic basis within a unit cell.
\item $N_b(j)$ is the number of basis matrices in the $j^{th}$ set.
\end{itemize}
The variables and quantities which we will be using are:
\begin{itemize}
\item ${\bf R}_{ji}$ : is the lattice vector corresponding to the
        $j^{th}$ triplet set and $i^{th}$ member.
\item $\tau_{ji},\tau'_{ji}$ : are the indices of the basis atoms
                             of the $j^{th}$ set and $i^{th}$ member.
\item ${\bf C}(j)$ : is a $3 \times 3$ matrix for
the $1^{st}$ member of
 the $j^{th}$ set.
It is the $\tau_{j1},\tau'_{j1}$ block of the IFCs in 
real-space with separation vector ${\bf R}_{j1}$, i.e.,
 ${\bf C}(j) = {\bf C}_{\tau_{j1},\tau'_{j1}}({\bf R}_{j1})$.
\item ${\bf F}_{jk}$ : for a given $j$, 
$\lbrace {\bf F}_{jk},k=1, \ldots, N_b(j) \rbrace$
is a linearly independent set
                   of $N_b(j)$ $3 \times 3$ matrices (the basis matrices). 
               They are related to ${\bf C}(j)$ by
\begin{equation}
{\bf C}(j) = \sum_{k=1}^{N_b(j)} a_{jk} {\bf F}_{jk},
\end{equation}
where the $a_{jk}$ are to be determined by a fitting procedure to be described and are
not determined by symmetry.
\item ${\bf C}(\mu,\sigma,\sigma')$ : is a $3 \times 3$ 
matrix giving the $\sigma,\sigma'$
          block in the IFCs at Bloch vector ${\bf k}_\mu$.
\item ${\bf U}_{ji}$ : is the matrix transformation that relates
               an element in the set/member $ji$ with an element
               with set/member $ji$, according to,
${\bf C}_{\tau_{ji} \tau'_{ji}}({\bf R}_{ji}) = {\bf U}_{ji}^T {\bf C}_{\tau_{j1}\tau'_{j1}}({\bf R}_{j1}) {\bf U}_{ji}$.
\end{itemize}
We consider the following quantity $s$, defined as,
\begin{equation}
s = \sum_{\sigma \sigma'}\sum_{ab\mu}
|\sum_{\lbrace i,j|\tau_{ji} = \sigma,\tau_{ji}' = \sigma'\rbrace} 
[{\bf U}^{T}_{ji} {\bf C}(j) {\bf U}_{ji}]_{ab} e^{i{\bf k}_\mu \cdot {\bf R}_{ji}} - 
[{\bf C}(\mu,\sigma,\sigma')]_{ab}
|^2 w(\mu),
\end{equation}
or
\begin{equation}
s(\lbrace a_{jk} \rbrace) = \sum_{\sigma \sigma'}\sum_{ab\mu}
|\sum_{\lbrace i,j,k|\tau_{ji} = \sigma,\tau_{ji}' = \sigma'\rbrace} 
[{\bf U}^{T}_{ji} {\bf F}_{jk} {\bf U}_{ji}]_{ab} a_{jk}
e^{i{\bf k}_\mu \cdot {\bf R}_{ji}} - 
[{\bf C}(\mu,\sigma,\sigma')]_{ab}
|^2 w(\mu),
\label{chisquared}
\end{equation}
where here we show the explicit dependence on the set of coefficients
$\lbrace a_{jk} \rbrace$ and the weights  $\lbrace w(\mu) \rbrace$, which are arbitrary and can be chosen for a variety of convenient ends, such as phase-space weighting.
The matrix,
${\bf C}(\mu,\sigma,\sigma')$,
is determined by means described in preceding sections of this paper.
The parameters $\lbrace a_{jk} \rbrace$ are fitting parameters.
Here 
we choose the triplet sets, $(j,i)$, for which we have
\[
\parallel {\bf R}_{ji} + {\bf \tau'}_{ji} -
                {\bf \tau}_{ji} \parallel  < r_c,
\]
where $r_c$ is some cutoff selected by the user.

\indent To ensure that
$\tilde{{\bf C}}_{\tau \tau'}({\bf k})$ is manifestly Hermitian, we
divide the triplet sets into two groups. 
Within group (1) are those for which there 
exists a paired counterpart in a different set, 
i.e., a pair  $(j,i)$ and $(j',i')$, 
such that we have ${\bf R}_{ji} = - {\bf R}_{j'i'}$ and
$\tau_{ji} = \tau'_{j'i'}, \tau'_{ji} = \tau_{j'i'}$.
Within group (2) are those
for which there exists no such pair.
We let 
$Q{\lbrace i,j|\tau_{ji} = \sigma,\tau_{ji}' = \sigma'\rbrace}$ denote
the set corresponding to group (2) and 
$P{\lbrace i,j|\tau_{ji} = \sigma,\tau_{ji}' = \sigma'\rbrace}$ denote
the set corresponding to group (1) with only one of the pair members
counted to avoid overcounting.  (Note that the triplet sets in group (1) occur in pairs.) 
This implies including half of the triplet sets in group (1) when performing
the fit, and using the Hermiticity condition to deduce
${\bf C}_{\tau \tau'}({\bf R})$ if it belongs to an excluded triplet set.
It should not matter which of a pair of triplet sets is included in the 
fit. 

\indent We then rewrite Eq.~\ref{chisquared}
in terms of these two groups as:
\begin{eqnarray*}
s(\lbrace a_{jk} \rbrace) &=& \sum_{\sigma \sigma'}\sum_{ab\mu}
|\sum_{Q\lbrace i,j,k|\tau_{ji} = \sigma,\tau_{ji}' = \sigma'\rbrace} 
a_{jk} [{\bf U}^{T}_{ji} {\bf F}_{jk} {\bf U}_{ji}]_{ab}
e^{i{\bf k}_\mu \cdot {\bf R}_{ji}}- 
[{\bf C}(\mu,\sigma,\sigma')]_{ab}
|^2 w(\mu)\\
&+& \sum_{\sigma \sigma'}\sum_{ab\mu}
|\sum_{P\lbrace i,j,k|\tau_{ji} = \sigma,\tau_{ji}' = \sigma'\rbrace} 
a_{jk} [{\bf U}^{T}_{ji} {\bf F}_{jk} {\bf U}_{ji}]_{ab} 
e^{i{\bf k}_\mu \cdot {\bf R}_{ji}} - 
[{\bf C}(\mu,\sigma,\sigma')]_{ab} |^2 w(\mu)\\
&+&  \sum_{\sigma \sigma'}\sum_{ab\mu}
|\sum_{P\lbrace i,j,k|\tau_{ji} = \sigma,\tau_{ji}' = \sigma'\rbrace} 
a_{jk}  [{\bf U}^{T}_{ji} {\bf F}_{jk} {\bf U}_{ji}]_{ba} 
e^{-i{\bf k}_\mu \cdot {\bf R}_{ji}} - 
[{\bf C}(\mu,\sigma',\sigma)]_{ab} |^2 w(\mu).
\end{eqnarray*}
We perform a constrained minimization
on the quantity $s(\lbrace a_{jk} \rbrace)$ subject to
the acoustic sum rule, which can be expressed by
\begin{equation}
\sum_{{\bf R} \tau'} [{\bf C}_{\tau \tau'}({\bf R})]_{ab} = 0.
\label{asr}
\end{equation}
Here there is one equation for every combination of $\tau$, $a$, and $b$,
giving a total of $9N$ constraint equations. The acoustic sum
rule is therefore a further 
set of constraint equations linear in the
fitting parameters. To determine these linear equations explicitly, 
we first define the
quantity, 
\begin{equation}
\end{equation}
\begin{eqnarray*}
A_{jk;\mu\sigma\sigma'ab} &=&
\sum_{Q \lbrace i|\tau_{ji} = \sigma,
\tau'_{ji} = \sigma' \rbrace }
[{\bf U}^{T}_{ji} {\bf F}_{jk} {\bf U}_{ji}]_{ab} 
e^{i {\bf k}_\mu \cdot {\bf R}_{ji}}
\\
&+& \sum_{P\lbrace i|\tau_{ji} = \sigma, \tau'_{ji} = \sigma' \rbrace }
 [{\bf U}^{T}_{ji} {\bf F}_{jk} {\bf U}_{ji}]_{ab} 
e^{i {\bf k}_\mu \cdot {\bf R}_{ji}}\\
&+&\sum_{P\lbrace i|\tau_{ji} = \sigma',\tau'_{ji} = \sigma \rbrace }
[{\bf U}^{T}_{ji} {\bf F}_{jk} {\bf U}_{ji}]_{ba} e^{-i {\bf k}_\mu \cdot {\bf R}_{ji}}.
\end{eqnarray*}
The acoustic sum rule applies to the wave vector 
\begin{equation}
{\bf k}_\mu=0.
\label{Gam}
\end{equation}
It is helpful to take the value of $\mu$ from Eq.~\ref{Gam}, and define:
\begin{equation}
C_{jk;\sigma \sigma' ab} = {\it A}_{jk;\mu\sigma \sigma' ab}.
\end{equation}
The acoustic sum rule is satisfied through the fitting parameters
which satisfy the following system of linear equations:
\begin{equation}
\sum_{jk}
a_{jk} C_{jk;\sigma \sigma 'ab} = 0.
\label{constraint}
\end{equation}
\indent This system of $9 N$ equations is not necessarily linearly independent.
For example, in a system like diamond, there is only one linearly independent
constraint equation in  a set of 18 linear equations. To reduce
Eq.~\ref{constraint} to an equivalent smaller set of 
linearly independent equations, we use Gauss-Jordan elimination. Supposing
that they reduce to a system of $N_c$ linearly independent equations,
we have
\begin{equation}
g^{(\nu)}(\lbrace a_{jk} \rbrace) \equiv
\sum_{jk}
a_{jk} c^{(\nu)}_{jk} = 0, \nu = 1,\ldots,N_c.
\label{c}
\end{equation}
By minimizing $s(\lbrace a_{jk} \rbrace)$ subject
to the constraint relation in Eq.~\ref{c}, we have
\begin{equation}
{   {\partial s(\lbrace a_{jk} \rbrace)} \over
    {\partial a^*_{j'k'} } }
+ \sum_{\nu=1}^{N_c} \lambda^{(\nu)} 
{{\partial [g^{(\nu)}(\lbrace a_{jk} \rbrace)]^*} \over
{\partial a^*_{j'k'}}}  = 0.
\label{min} 
\end{equation}
Here $\lambda^{(\nu)}$ are Lagrange multipliers, which need to be found.
Eq.~\ref{min} reduces to
\begin{equation}
\sum_{jk} M_{j'k',jk} a_{jk} + 
\sum_{\nu=1}^{N_c} \lambda^{(\nu)} c^{(\nu)*}_{j'k'} = b_{j'k'},
\label{Ax=b}
\end{equation}
with
\begin{equation}
M_{j'k',jk} = \sum_{\mu \sigma \sigma' ab} w(\mu) 
{\it A}_{jk;\mu\sigma\sigma'ab}
{\it A}^*_{j'k';\mu\sigma\sigma'ab}
\end{equation}
and
\begin{equation}
b_{jk} = \sum_{\mu \sigma \sigma' ab} w(\mu)
{\it A}^*_{jk;\mu\sigma\sigma'ab}
[{\bf C}(n,\sigma,\sigma')]_{ab}.
\end{equation}
If there are $N_{f}$ fitting parameters (i.e., the $a$'s),
then Eqs.~\ref{c} and \ref{Ax=b} form a system of $N_c+N_{f}$ equations with
$N_c + N_{f}$ unknowns.

\begin{center}
\Large{IV.  Macroscopic Fields and Polar Crystals}
\end{center}
\indent For nonpolar crystals, and hence those with no optically active phonons, the method described above is sufficient for the calculation of full phonon dispersion.  However, polar materials, in which distortions interior to the primitive cell can generate electric dipoles, require further consideration.  This is because the boundary conditions of DFT calculations will not allow any physics associated with macroscopic electric polarizations or fields.  The latter are connected to the LO modes, and the DFT calculations are relevant only to the acoustic and TO phonon branches.  The macroscopic fields can stiffen the small-wave-vector, LO modes in a manner that can depend on the direction from which the wave vector approaches zero~\cite{Par98}.

\indent  The situation is further complicated by the fact that it is only at the BZ center that supercell DFT calculations do not include all the lattice-field interactions.  So the fitting procedure must reintroduce these interactions at the BZ center while leaving the dynamical matrices at other points unaffected.

\indent  To do this, the Born effective charges are first calculated~\cite{KSV}. These charges are defined as 
\begin{displaymath}
{Z^{*}}_{{\tau}{a}{b}}=\Omega\frac{{\partial}P_{a}}{{{\partial}u_{{\tau}b}}},
\end{displaymath} 
where $\bf {P}$ is the macroscopic polarization (in units of dipole per unit volume) and ${\bf{u}}_{\tau}$ is a sublattice shift.  These charges must be zero for a nonpolar crystal, and where they are zero there is no energy difference between the LO and TO modes in the long-wave limit.  

\indent  Considering two ions in the basis $\tau$ and $\tau'$  at sites separated by lattice vector ${\bf R}$, the contribution to the real-space IFCs from the screened electrostatic energy between the two is 
\begin{equation}
{C^{Coul}_{\tau\tau'ab}({\bf{R}})}=\frac{{{{\partial}^{2}E^{Coul}}}}
{{\partial}{{x}}_{\tau{a}}({\bf R}'){\partial}{{x}}_{\tau'{b}}({\bf R}'+{\bf 
R})}.
\label{cforce}
\end{equation}
The Coulomb energy here is associated with dipole-dipole interactions, and is given by:
\begin{displaymath}
E^{Coul}=
\frac{1}{2}\sum_{{\bf R}'{\bf R}{\tau}{\tau}'{b}{b}'{a}{a}'}\frac{{Z^{*}_{{\tau}{a}{b}}}
{Z^{*}_{{\tau}'{a'}{b'}}}}
{\epsilon_{\infty}}
\left(
\frac{{\delta}_{aa'}}{d^3}
-3\frac{d_{a}d_{a'}}
{d^5}
\right){{x}}_{\tau{b}}({\bf R}){{x}}_{{\tau}'{b'}}({\bf R}').
\end{displaymath}
The above should be understood to exclude terms corresponding to the same ion on the same site.  The relative coordinate is indicated by ${\bf d}$
\begin{displaymath}
{\bf d}={\bf R}+{\bf s}_{\tau{'}}-{\bf s}_{{\tau}},
\end{displaymath}
 and $\epsilon_{\infty}$ is the dielectric constant.
The derivatives are with respect to displacements from equilibrium along the corresponding ionic coordinates.   Gonze and Lee~\cite{Gonze} and Shirley \textit{et al.}~\cite{Shirley} write in detail the reciprocal-space transform of Eq.~\ref{cforce}, 
\begin{equation}
\tilde{{\bf C}}^{Coul}({\bf k})=\sum_{\bf R}{\bf C}^{Coul}({\bf R})e^{i{\bf k}\cdot{\bf R}}
\label{etrans}
\end{equation}
in accord with the Ewald formulation~\cite{Ewald}.
Eq.~\ref{etrans} can be written as:
\begin{equation}
\tilde{\bf C}^{Coul}({\bf k})=\tilde{\bf C}^{na}({\bf k})+\tilde{\bf C}^{an}({\bf k}),
\end{equation}
where the analytical and nonanalytical contributions are denoted with superscripts.  The nonanalytic contribution arises from the ${\bf G}=0$ term in the reciprocal-lattice Ewald sum, and accounts entirely for the LO-TO splitting. Its limiting behavior for small wave vector can be shown to be~\cite{smallk}:
\begin{displaymath}
{{\tilde{C}}^{na}}_{{\tau}{\tau}'bb'}({\bf k}{\rightarrow}0)=\frac{4{\pi}}{{{\epsilon}_{\infty}}\Omega}
\frac
{{\sum}_{aa'}{Z^{*}}_{{\tau}{ab}}{Z^{*}}_{{\tau}'{a'b'}}k_{a}k_{a'}}
{{\sum}_{aa'}k_{a}k_{a'}}
.
\end{displaymath}

\indent  The small-$\bf{k}$, nonanalytic contribution to $\tilde{{\bf C}}^{Coul}(\bf{k})$ is missing in a supercell calculation.  In order to reintroduce it, 
Eq.~\ref{identity}
can be modified as:
\begin{equation}
\tilde{{\bf C}}^{DFT}({\bf{k}})-{{\bf T}}({\bf{k}})=\sum_{\bf{R}}
e^{i{\bf{k}}\cdot{\bf{R}}}
\hat{\bf C}({\bf{R}}),
\label{5}
\end{equation}
where $\tilde{\bf C}^{DFT}({\bf{k}})$ indicates the matrix calculated within DFT and excludes the effects of the macroscopic electric field, and ${\bf T}({\bf k})$ is defined as:
\begin{equation}
{{\bf T}}({\bf{k}})=
\left\{\begin{array}{cc}
&{{\tilde{\bf C}^{an}}}({\bf{k}}):
{\bf{k}}=0\\
&{{\tilde{\bf C}^{Coul}}}({\bf k}):
{\bf{k}}{\neq}0.
\end{array}
\right.
\end{equation}

The distinction between the two cases is always clear, because we are only considering the sampled BZ points, and all of these points, except for the ${\Gamma}$ point itself, are far from the BZ center.
Within the fitting procedure, the real-space and reciprocal-space IFCs are now replaced by $\hat{{\bf C}}({\bf{R}})$, and $\tilde{\bf C}^{DFT}({\bf k})-{{\bf T}}({\bf{k}})$, respectively.

\indent Then, the full reciprocal-space IFC matrix, with the contribution of the macroscopic fields, is:
\begin{equation}
{{\tilde{\bf C}}({\bf{k}})}
=
\tilde{\bf C}^{Coul}({\bf k})+
{\sum_{\bf{R}}}
\hat{{\bf C}}({\bf{R}})e^{i{\bf{k}}\cdot{\bf{R}}}.
\label{8}
\end{equation}
\begin{center}
\Large{V. Computational Details}
\end{center}

\begin{figure}
\begin{center}
\includegraphics[scale=0.5,angle=-90]{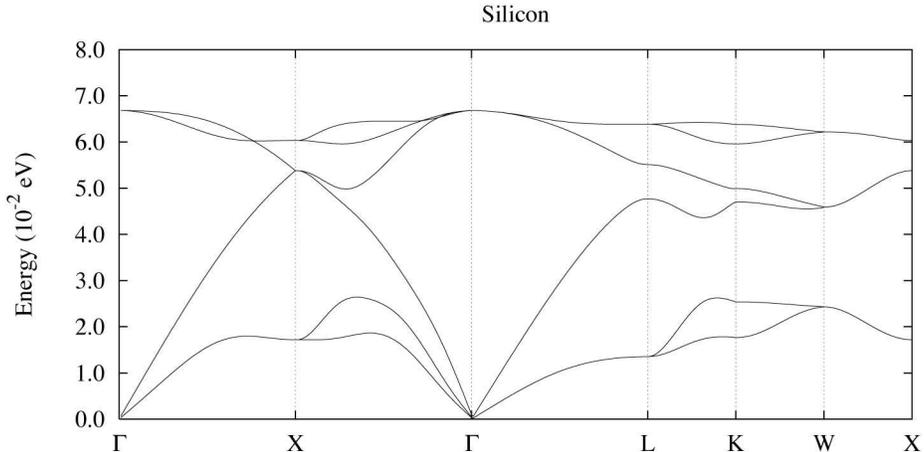}
\caption{\label{fig:1}Phonon dispersion in Si.}
\end{center}
\end{figure}

\begin{figure}
\begin{center}
\includegraphics[scale=0.5,angle=-90]{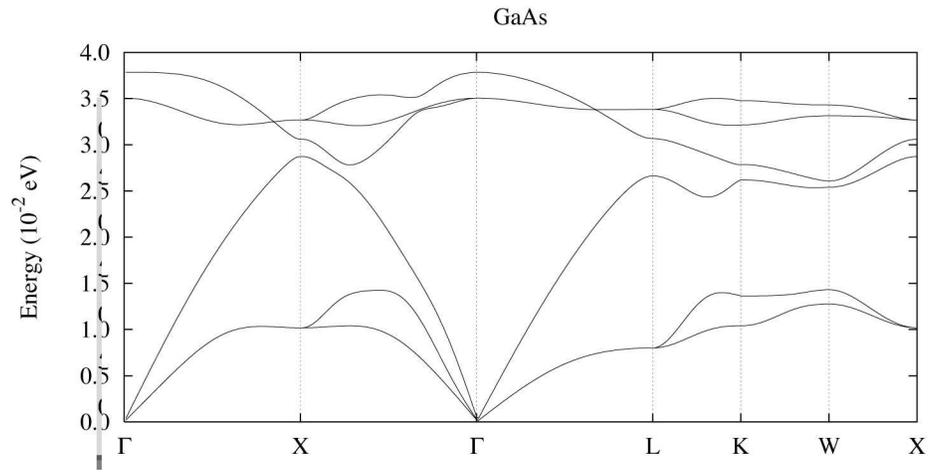}
\caption{\label{fig:2}Phonon dispersion in GaAs.}
\end{center}
\end{figure}

\begin{figure}
\begin{center}
\includegraphics[scale=0.5,angle=-90]{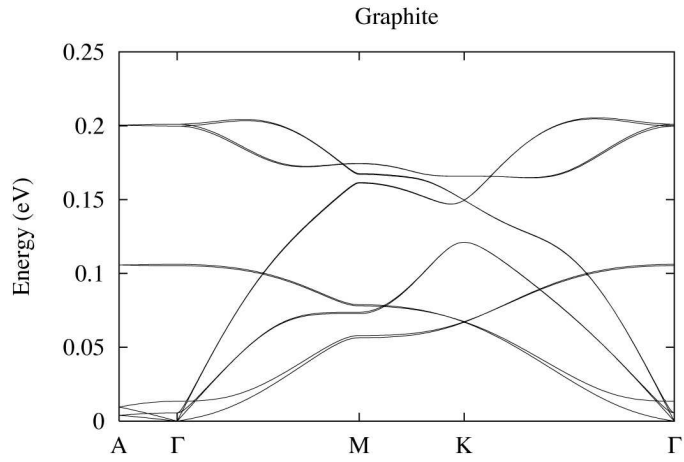}
\caption{\label{fig:3}Phonon dispersion in graphite.}
\end{center}
\end{figure}

\indent Dispersion curves for Si, GaAs and graphite are presented in Figs. 
\ref{fig:1},\ref{fig:2},\ref{fig:3}.  Below are listed which phonon ${\bf k}$ points are used in each fit, and which real-space triplet matrices are included.  These particular selections are somewhat arbitrary, so long as the BZ is adequately sampled and the triplet matrices are overdetermined.  For the case of graphite, extra weight is given to the ${\bf \Gamma}$ point in the fit to ensure that at very small ${\bf k}$ the very small eigenvalues of the dynamical matrix, ${\omega}^2$, do not go negative.

\indent  The symmetries are determined from lattice structure:  for GaAs and Si, the lattice vectors are
\begin{displaymath}
{\bf a}_1=\frac{a}{2}(0,1,1),~{\bf a}_2=\frac{a}{2}(1,0,1),~{\bf a}_3=\frac{a}{2}(1,1,0).  
\end{displaymath}
A lattice constant of 10.05 $a_0$ is employed for Si, and 10.46 $a_0$ for GaAs (all reported lattice constants correspond to the values for the theoretically relaxed crystals).  The basis vectors for each crystal are:
\begin{displaymath}
{\bf s}_{{\tau}_1}=-\frac{1}{8}({\bf a}_1+{\bf a}_2+{\bf a}_3),
~{\bf s}_{{\tau}_2}=\frac{1}{8}({\bf a}_1+{\bf a}_2+{\bf a}_3). 
\end{displaymath}
In Si, the two basis sites are symmetry-mapped, and in GaAs our calculation identifies the Ga with ${\tau}_1$.

\indent  The lattice vectors of graphite are:
\begin{displaymath}
{\bf a}_1={a}\left(\frac{\sqrt{3}}{2},\frac{1}{2},0\right),{\bf a}_1={a}\left(\frac{\sqrt{3}}{2},-\frac{1}{2},0\right),{\bf a}_3=c(0,0,1),  
\end{displaymath}
the in-plane lattice constant is 4.88 $a_0$, and the lattice parameters take the ratio, $c/a=2.72$.
The basis for graphite is
\begin{displaymath}
{\bf s}_{{\tau}_1}=\frac{1}{4}{\bf a}_3, 
~{\bf s}_{{\tau}_2}=-\frac{1}{4}{\bf a}_3,
~{\bf s}_{{\tau}_3}=-\frac{1}{3}\left({\bf a}_1+{\bf a}_2\right)+
\frac{1}{4}{\bf a}_3,
~{\bf s}_{{\tau}_4}=\frac{1}{3}\left({\bf a}_1+{\bf a}_2\right)-
\frac{1}{4}{\bf a}_3.
\end{displaymath}

\indent  For GaAs, the ${\bf k}$ points included in the fit are ${\bf \Gamma},~{\bf X},~{\bf L},~{\bf K},~\frac{2}{3}{\bf X},~\frac{2}{3}{\bf L},~\frac{1}{2}{\bf X},$
$~\frac{1}{2}{\bf L},~\frac{3}{4}{\bf K}, {\bf W}$, using cubic convention, and the lower-symmetry point $\frac{2\pi}{a}(\frac{3}{4},\frac{3}{4},\frac{-1}{4})$.  The prototypical triplets are the basis pairs 
$({\tau}_1{\tau}_1),({\tau}_1{\tau}_2),$ and $({\tau}_2{\tau}_2)~$
for ${\bf R}=(0,0,0)$; 
the basis pairs $({\tau}_1{\tau}_1),({\tau}_1{\tau}_2),$ and $({\tau}_2{\tau}_2)~$ 
for ${\bf R}=\frac{a}{2}(1,1,0)$; and the basis pairs 
$({\tau}_1{\tau}_1)$ and $({\tau}_2{\tau}_2)$ for ${\bf R}={a}(1,0,0)$.  For Si the DFT calculated phonons are the same set as for GaAs. 
The triplets are pairs 
$({\tau}_1{\tau}_1)$ and $({\tau}_1{\tau}_2)$
for ${\bf R}=(0,0,0)$; 
$({\tau}_1{\tau}_1)$ and $({\tau}_1{\tau}_2)$
for ${\bf R}={\bf a}_1$; and
$({\tau}_1{\tau}_1)$
for ${\bf R}={\bf a}_1+{\bf a}_2-{\bf a}_3$.
All triplets symmetry-related to those within the prototypical set enter the computation. 

\indent  The graphite calculation includes the phonons at
${\bf \Gamma}$,
${\bf K}$,
${\bf M}$,
${\bf A}$,
${\bf W}$,
$\frac{2}{3}{\bf M}$,
$\frac{1}{2}{\bf M}$,
 and 
$\frac{3}{4}{\bf K}.$
The graphite triplets are:
the pairs 
${(\tau}_1{\tau}_1),({\tau}_1{\tau}_2),({\tau}_1{\tau}_3),$
$({\tau}_1{\tau}_4),({\tau}_3{\tau}_3),$ and $({\tau}_3{\tau}_4)$ 
for ${\bf R}=(0,0,0);$
the pairs
${(\tau}_1{\tau}_1),({\tau}_1{\tau}_2),({\tau}_1{\tau}_3),({\tau}_3{\tau}_3),$ and $({\tau}_3{\tau}_4)$ 
for ${\bf R}=-{\bf a}_1$; 
the pairs 
$({\tau}_1{\tau}_3)$ and $({\tau}_2{\tau}_4)$
for ${\bf R}={\bf a}_1-{\bf a}_2$; and
the pairs $({\tau}_1{\tau}_1),({\tau}_2{\tau}_2),({\tau}_3{\tau}_3),$ and $({\tau}_4{\tau}_4)$
for ${\bf R}={\bf a}_2-2{\bf a}_1$.
\begin{center}
\Large{VI.  Conclusions}
\end{center}

\indent    We have made some innovations within the direct approach to the computation of lattice dynamics.  These include the calculation of dynamical matrices without resort to real-space atomic force constants, which may reduce the supercell size necessary for a full dispersion calculation; the general extension to polar crystals; and a symmetry-preserving fit algorithm for interpolation throughout the BZ.  The direct method is shown to render a complete, first-principles characterization of the phonon spectrum for a range of materials.  Accordingly, the direct method can be considered complimentary to density-functional perturbation theory as a means of computing a variety of fundamental, phonon-related solid-state properties.

\newpage

\end{document}